\documentclass[aps,prb,twocolumn]{revtex4}  
\usepackage{graphicx}
\usepackage{amssymb}                

\def\gapx{\lower 2pt \hbox{$\buildrel>\over{\scriptstyle{\sim}}$\ }}
\def\lapx{\lower 2pt \hbox{$\buildrel<\over{\scriptstyle{\sim}}$\ }}
\def\he4{$^4$He}
\def\paraH2{{\it p}-H$_2$}
\def\orthoD2{{\it o}-D$_2$}
\def\Am2{\AA$^{-2}$}

\begin{document}

\title{Enhanced Superfluid Response of Parahydrogen in Nanoscale Confinement}
\author{Tokunbo Omiyinka and Massimo Boninsegni} 
\affiliation{Department of Physics, University of Alberta, Edmonton, 
    Alberta, Canada T6G 2E7}
\date{\today}

\begin{abstract} 
Confinement has generally the effect of suppressing order in condensed matter. Indeed, phase transitions such as freezing, or the superfluid transition in liquid helium, occur at lower temperatures in confinement than they do in the bulk. 
We provide here an illustration of a physical setting in which the opposite takes place.
Specifically, the enhancement of the superfluid response of parahydrogen confined to nanoscale size cavities
is demonstrated by means of first principle computer simulations. Prospects to stabilize and observe the long sought but 
yet elusive bulk superfluid phase of parahydrogen in purposefully designed porous media are discussed.

\end{abstract}

\maketitle
\section{Introduction}
Helium is the only known element to turn superfluid at low temperature, in its equilibrium liquid phase. Among naturally occurring substances, a
potential second superfluid is parahydrogen (\paraH2), whose elementary constituents are molecules of spin zero (thus  obeying Bose statistics) and of mass approximately one half of that of a helium atom.  It was first suggested in 1972   that a fluid of  \paraH2 molecules should undergo a superfluid transition at a temperature close to 6 K.\cite{ginzburg}
However, the observation of this hypothetical superfluid has so far been prevented by the simple fact that, unlike helium, \paraH2 solidifies \cite{note} at a temperature of 13.8 K, under the pressure of its own vapour. This is due to the depth of the attractive well of the interaction between two \paraH2 molecules, about three times greater than that between two helium atoms, imparting to the system a strong propensity to crystallize, even in reduced dimensions.\cite{me} 
\\ \indent
It is experimentally known that the freezing temperature of a fluid can be substantially lowered from its bulk value, by confining it in a porous medium, such as vycor glass,\cite{tell,molz} which can be thought of as a random network of interconnected cavities of average size around 4 nm.\cite{mason} However, attempts to supercool liquid hydrogen by embedding it in a vycor matrix failed to yield any evidence of 
superfluid behaviour.\cite{bretz81,schindler96}  Indeed, neutron scattering studies of \paraH2 in vycor glass at low temperature suggest that the system crystallizes in the pores, albeit with a different crystal structure \cite{sokol}, a fact that can be attributed to the irregular geometry as well as to the strong attraction exerted on the \paraH2 molecules by the pore surface, where the crystal phase nucleates. It should also be mentioned that experimentally  one observes the lowering of the superfluid transition temperature of liquid helium confined in vycor,\cite{chan} to indicate that, while stabilizing the liquid phase, confinement has also the effect of suppressing superfluidity. 
\\ \indent
The questions remain open of whether a different confining environment may  make it feasible  to stabilize a superfluid phase of \paraH2, and of how such a confining medium would differ from vycor.
A hint may come from quantitative theoretical predictions that have been made \cite {sindzingre,kwon,noi} of superfluid behaviour in small \paraH2 clusters (thirty molecules or less), at temperatures of the order of a fraction of a K, of which some experimental confirmation has been obtained \cite{grebenev} (albeit perhaps not definitive \cite{callegari}).  The largest clusters for which  a significant superfluid response is predicted at a temperature $T \simeq$ 1 K, have a size of approximately 1.6 nm, and comprise $N$=27 \paraH2 molecule;\cite{noi} superfluidity is suppressed  in clusters of greater size, as crystalline order, originating at the center, becomes predominant. This suggests that a porous medium with cavities around 2 nm in size may prove suitable; such a medium ought to be a weaker adsorbent than silica, so as to suppress crystallization of \paraH2 at the pore surface. 
\\ \indent
In order to explore the above scenario and to gain insight in the effect of confinement on the superfluid response of \paraH2, we have  carried out first principle computer simulations of a microscopic model of \paraH2 in confinement. The main physical conclusions expounded here is that nanoscale confinement can greatly {\it enhance} the superfluid response of parahydrogen.
\section{Model}
We consider a system of $N$ \paraH2 molecules, described as point particles of spin zero, enclosed in a spherical cavity of radius $R$, described by the following quantum-mechanical many-body Hamiltonian:
\begin{equation}
\label{ham}
\hat H = -\lambda\sum_i\nabla^2_i +\sum_{i<j}v(r_{ij})+\sum_i V(r_i).
\end{equation}
Here,  ${\bf r}_i$ are the positions of all \paraH2 molecules, measured with respect to the center of the cavity (set as the origin), $r_{ij}\equiv |{\bf r}_i-{\bf r}_j|$, $\lambda=12.031$ K\AA$^2$, $v$ describes the interaction of a pair of molecules, whereas $V$ that of each molecule with the cavity. For $v$, which we regard as spherically symmetric, we use both 
the well-known pair potential of Silvera and Goldman,\cite {SG} as well as
a recently proposed modification \cite{moraldi,tokunbo} thereof,  which has been shown to afford an accurate reproduction of the experimental equation of state of solid \paraH2 up to megabar pressure. Computed structural and energetic properties, as well as the main conclusion regarding the superfluid response, do not depend on which potential is used.
\\ \indent
For $V$, assumed to depend only on the distance of a molecule from the center of the cavity, we take the following model potential:\cite{gatica}
\begin{equation}\label{potl}
V(r)=2D \biggl \{\frac{b^9 F(x)}{(1-x^2)^9}-\frac{6b^3}{(1-x^2)^3}\biggr\},
\end{equation}
where $x\equiv r/R$, $F(x)=5+45x^2+63x^4+15x^6$, $b\equiv(a/R)$ and $a$ and $D$ are two parameters that are adjusted to reproduce, as closely as allowed by such a relatively crude model, a specific particle-substrate interaction. 
Expression (\ref{potl}) is the extension to the case of a spherical cavity of the so-called ``3-9" potential, describing the interaction of a  particle with an infinite, planar substrate.
Specifically, $D$ has the dimensions of an energy, and is  essentially the depth of the attractive well of the potential experienced by the particle in the vicinity of the substrate; on the other hand, $a$ is a characteristic length, roughly speaking the distance of closest approach to the substrate, where the molecule begins to experience a strong repulsion.
\\ \indent
The model of the system utilized here is based on the physical picture of bulk \paraH2 in a porous medium as ``broken down" in loosely interconnected  clusters, each cluster enclosed in a tight environment, described here as a spherical cavity. Such a geometry was chosen for its greater simplicity over others (e.g., cylindrical) which might be regarded as more realistically describing existing porous materials, but whose study would require the introduction of additional parameters (e.g., the characteristic length of a cylinder), in turn complicating the inference of the basic physics.
The model clearly contains other major simplifications, notably the assumption that the interaction between hydrogen molecules can be sufficiently accurately described by a spherically symmetric pair potential, as well as the fact that the cavity is regarded as perfectly spherical and smooth. Although one might expect that the interaction between two \paraH2 molecules be significantly altered inside a nanoscale cavity, the weakness of the substrate seems to justify the use of a central pair potential, based on the expected minimum distance from the surface (over 3 \AA) at which molecules will be sitting (this is confirmed by our results). The same argument justifies the neglect of surface defects and corrugation, an assumption routinely made in numerical studies of adsorption of He or \paraH2 on alkali substrates.\cite{io}
\\ \indent
More generally, despite its simplicity such a model allows us to address the physical question that we wish to pose here, namely the effect of confinement on the superfluid response of \paraH2. Equivalent, or even simpler models (e.g., cavities with hard walls) have been utilized to study structure of  $^4$He and classical fluids in confinement.\cite{delma,chui}
\section{Methodology}
We have carried out a theoretical investigation of the low temperature physical properties of the system described by Eqs. \ref {ham} and \ref{potl}, by means of first principle computer simulations. Specifically, we used the Worm
Algorithm in the continuous-space path integral representation. Because this well-established computational methodology is thoroughly described elsewhere, we shall not review it here.\cite{worm,worm2} The most important aspects to be emphasized here, are that it enables one to compute thermodynamic properties of Bose systems at finite temperature, directly from the microscopic Hamiltonian, and that it grants direct 
access to  energetic, structural and superfluid properties of the confined \paraH2 fluid, in practice with no approximation.
\\ \indent
We illustrate below two sets of simulation results, both obtained by setting the cavity radius $R$ to 10 \AA, but with two distinct choices of the parameters $D$, $a$, corresponding to very different adsorption strengths. The first choice, henceforth labeled with Cs, has $D$=37.82 K and $a$=3.88 \AA; these are the  recommended \cite{chi} values to describe the interaction of a \paraH2 molecule with a Cs substrate, one of the most weakly attractive known. Its weakness is such that, as experiments show, a hydrogen fluid fails to wet it at low temperature,\cite{ross} as computer simulations also show.\cite{io}

The second choice, namely $D$=100 K and $a$=2.05 \AA, is roughly in the ballpark of what one would expect for \paraH2 molecules near a silica substrate;\cite{treiner} thus, we shall henceforth refer to the scenario described 
by this parameter set as 
Glass. It should be stressed, however, that in neither case do we aim at reproducing accurately 
any realistic interaction (which requires more elaborate 
functional forms anyway), as that turns out not to matter much in the end, as we show below.
Rather, our aim is that of investigating opposite ends of the adsorption continuum.  The considerably greater well depth and shorter range of the repulsive core, render the ``Glass" much more attractive to \paraH2 molecules than the Cs cavity.
\section{Results}
\subsection{Energetics}
\begin{figure}  [t]           
\centerline{\includegraphics[scale=0.34]{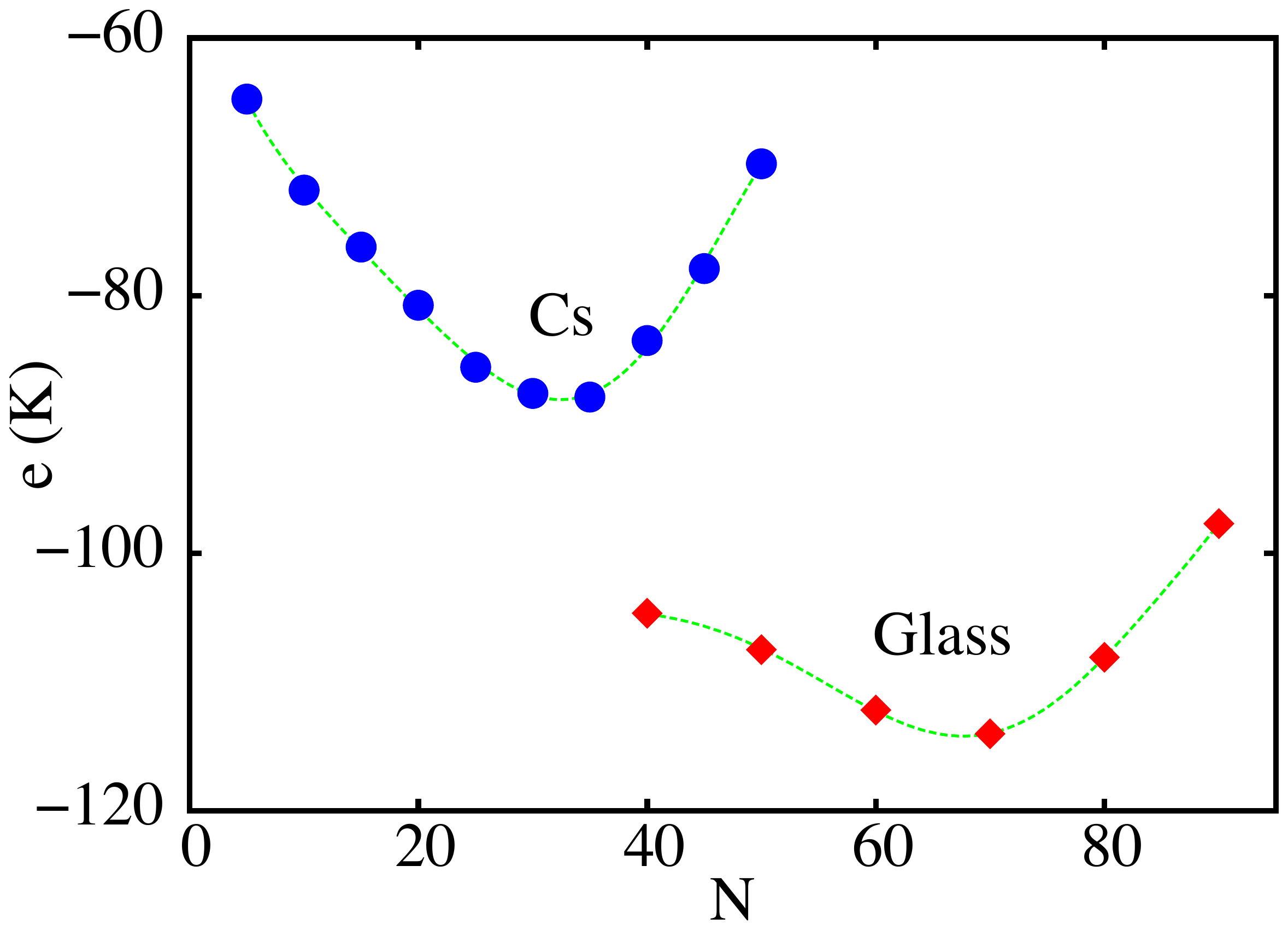}} 
\caption{Energy per hydrogen molecule $e$ (in K) versus number $N$ in the $T\to 0$ limit, inside
a Cs (filled circles) and Glass (diamonds) cavities of radius 10 \AA. Dashed lines are fits to the data. 
Statistical errors are at the most equal to symbol size.}
\label{ene}
\end{figure}

We begin the illustration of the results of the simulations by discussing the computed energetics. We obtain ground state estimates by extrapolating to $T=0$ low temperature results. In pratice, we find that
energy values, as well as radial density profiles, remain unchanged, on the scales of the figures shown here, below 
$T \lesssim$ 4 K. Fig. \ref{ene} displays the energy per \paraH2 molecule (in K) as a function of the number of molecules in the cavity. Both curves feature minima at specific numbers of molecules, which correspond to the minimum filling of the cavity, at thermodynamic equilibrium. The curve for Glass is shifted some 30 K downward with respect to that for Cs, and its minimum attained for a value of $N$ close to seventy molecules, over twice as much as the corresponding one for a Cs cavity. This is of course consistent with the greater adsorption exerted by the Glass cavity, and gives us an idea of the range within which $N$ can vary, inside a cavity of this size. An interesting thing to note is that for a Cs cavity the minimum occurs at an energy close to $-88.5$ K, identical to the ground state chemical potential \cite{tokunbo} for solid \paraH2. This means that a cavity of this size is barely at the wetting threshold for a weekly adsorbing substrate such as Cs. 
\subsection{Structure}
\begin{figure}  [t]           
\centerline{\includegraphics[scale=0.36]{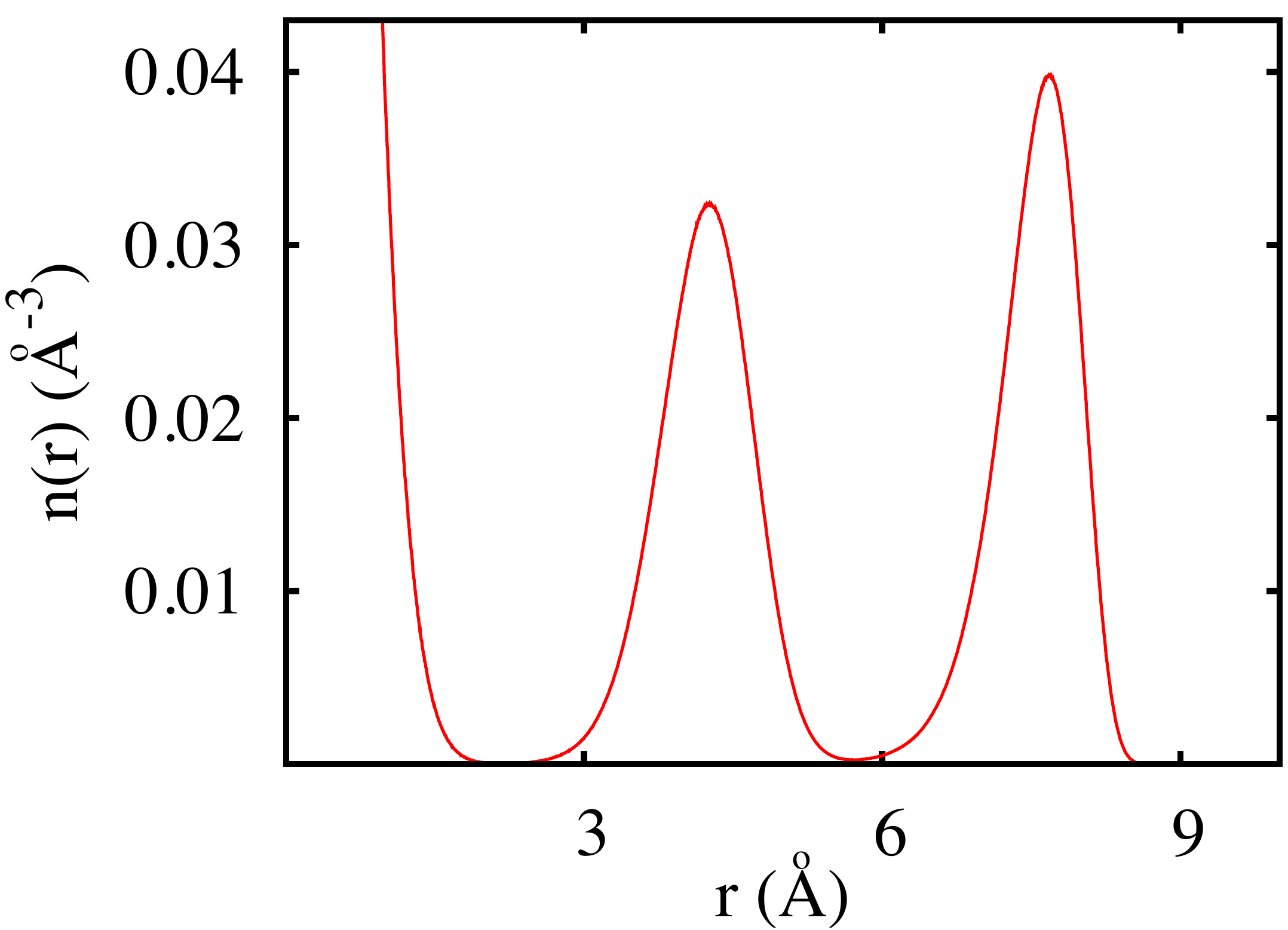}} 
\caption{Radial density profile $n(r)$ in (\AA$^{-3}$) for 70 \paraH2 molecules inside a Glass cavity at  $T=0.5$ K, modeled as explained in the text. The strong peak at the origin signals the presence of a paticle at the center of the cavity. Statistical errors are not visible on the scale of the figure.}
\label{radg}
\bigskip
\centerline{\includegraphics[scale=0.38]{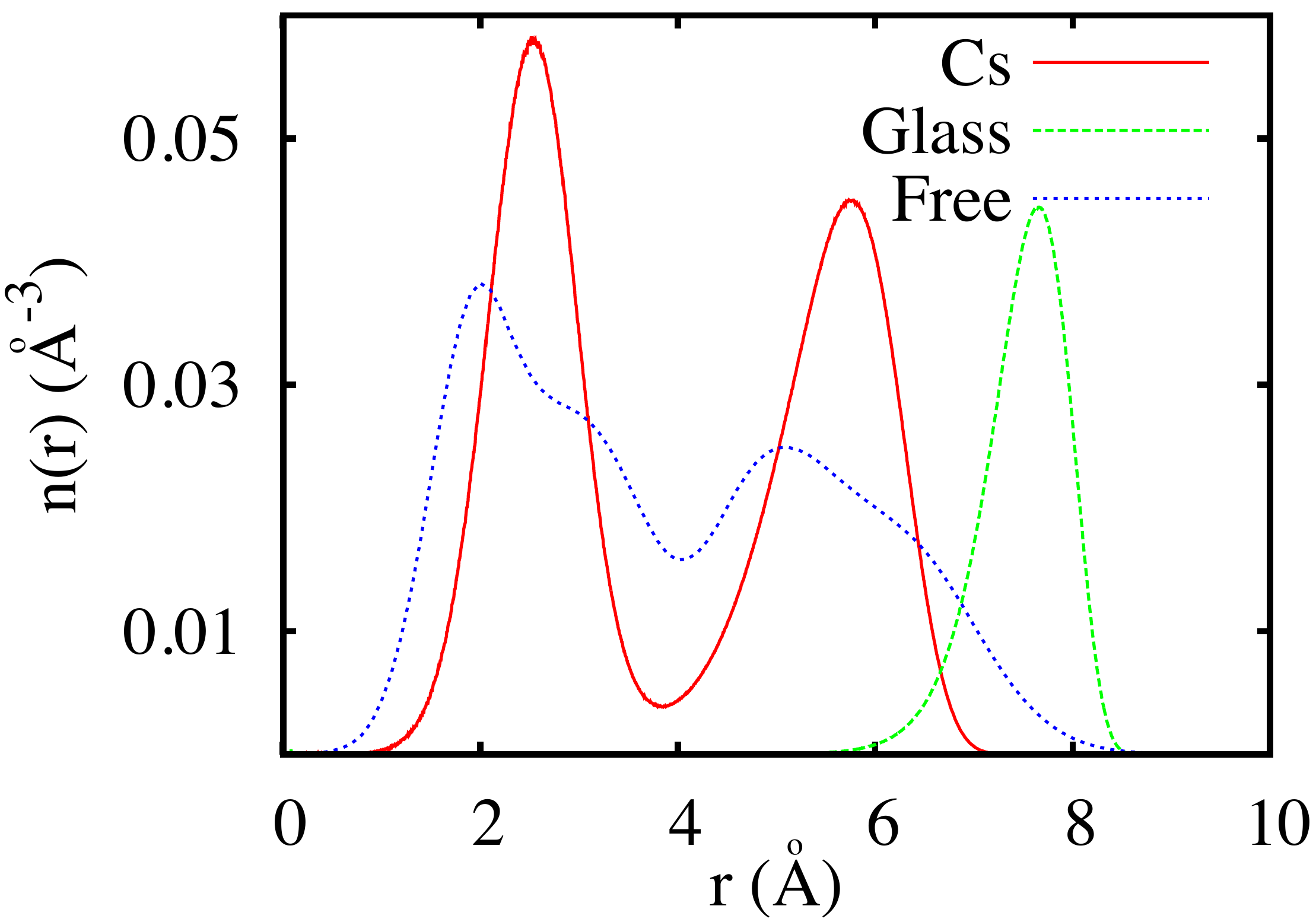}} 
\caption{Radial density profile $n(r)$ in (\AA$^{-3}$) for 32 \paraH2 molecules inside a Cs cavity (solid line) at $T=1$ K, modeled as explained in the text. Also shown are the profiles for the same number of molecules inside a Glass cavity (dashed line) and in a free standing \paraH2 cluster (dotted line). Statistical errors are small  on the scale of the figure.}
\label{radc}
\end{figure}

In order to gain insight on the structure of  \paraH2 inside the cavity, we examine the computed spherically averaged
radial density of molecules $n(r)$, as a function of the distance from the center of the cavity.  Fig. \ref{radg} shows this quantity at T=0.5 K for the case of seventy molecules inside the Glass cavity.  Density profiles for systems of sixty and fifty molecules look very similar, the heights of the two peaks shown in Fig. \ref{radg} being slightly reduced. 
The density profile shows that 
molecules are arranged on concentric spherical shells, corresponding to the sharp density peaks. The half width at half maximum of these peaks is less than 1 \AA, i.e., shells are rigid, molecular excursions in the radial direction being fairly limited.  This is true even within the shells, as molecules are scarcely mobile, held in place by the hard core repulsion of the intermolecular potential, which dominates the physics of the system at such dense packing.
Consequently, quantum-mechanical exchanges of molecules, which underlie the superfluid response in a quantum many-body system of indistinguishable particles, are strongly suppressed, and no appreciable superfluid response is observed in this case. 
\\ \indent
A very different physical scenario takes place in the Cs enclosure, as shown in Fig. \ref{radc}. Solid line shows the density profile for a cluster of 32 molecules,  i.e., the minimum of the energy curve in Fig. \ref{ene}, at $T=1$ K. Also shown (dashed line) is the density profile for the same number of molecules in a Glass cavity. There is a clear difference between the arrangements of the \paraH2 molecules in the various cases; in the Glass cavity, all molecules are close to the surface, whereas in the Cs one they form two concentric shells, the outer sitting considerably closer to the center of the cavity, and further away from its surface. This is, obviously, a consequence of the weakness of the Cs substrate compared to the Glass one, and of the consequent greater importance played by the repulsive core of the intermolecular potential at short distances.
\\ \indent
Another important feature is that, unlike in the case shown in Fig. \ref{radg}, the demarcation between the two shells present in the Cs cavity is not nearly as sharp as in the Glass one, as the density dips but does not go all the way to zero between the two peaks, as in Fig. \ref{radg}. This is suggestive of molecular delocalization, as well as of the possibility of significant quantum-mechanical exchanges and ensuing superfluidity.  In this respect, the density profile of 32 \paraH2 molecules in the Cs cavity is  closer to that of a free \paraH2 cluster comprising the same number of molecules (also shown in Fig. \ref{radc}, dotted line, computed separately in this work), than to that predicted in the stronger Glass enclosure. The profile of a free standing cluster extends further out, as no repulsive cavity wall is present. 
\subsection{Superfluidity}
The most important result of this study, however, is that the superfluid response at $T=1$ K of the cluster which is enclosed in the Cs cavity  is close to 50\%, whereas that of a free cluster at the same temperature is essentially zero (statistical noise level) \cite{area}. In other words,  {\em confinement has the effect of  greatly enhancing the superfluid response of the \paraH2 cluster}. This is confirmed by the occurrence of cycles including as many as 25 of the 32 molecules in the cluster that is enclosed in the Cs cavity, whereas no cycles comprising more than six molecules are observed in a free cluster.  The radial density profile of the free cluster displays considerably more structure in the inner shell, suggesting that molecules are more localized than inside the cavity.
If the cavity is made much more attractive, as is the case for the Glass one, exchanges are strongly suppressed, as molecules arrange orderly, to form a thin shell near the surface; on the other hand, in a weakly attractive cavity, confinement  renders the cluster more compact than in free space, increasing the propensity of molecules to exchange.
Thus, the environment experienced by molecules inside the  weak Cs cavity is such that the formation of a solid-like cluster, which takes place in free space, is frustrated; the cluster remains liquid-like, as the substrate is not attractive enough to promote crystallization near the surface, which is what happens in a strong adsorbents like silica or carbon.
\\ \indent
The simulations carried out here show that it is possible to enhance dramatically the superfluid response of the system in nanoscale size confinement, if the medium is weakly attractive.  The size of the cavity for which results were presented here (20 \AA\ diameter) seem to offer the right interplay of surface and bulk properties to lead to a novel phase; we also carried out simulations for Cs cavities of diameter 15 and 10 \AA, and found equilibrium superfluid clusters of around 17 molecules in the first case, no adsorption in the latter. If the diameter is made greater, bulk physics (i.e., the equilibrium crystalline phase of \paraH2) quickly becomes dominant and the superfluid response consequently is suppressed.
\section{Conclusions}
We can now discuss the implications of this study on the possible observation of the yet elusive bulk superfluid phase of \paraH2. The first thing that we need to assess, is the dependence of our results on the specific choice of geometry adopted in this study, namely that of a spherical cavity. The main physical effect observed here, that leads to the enhancement of intermolecular quantum exchanges, hinges on the weakness of the substrate and on the characteristic size of the confining region. Thus, we predict that it should also take place in different geometries (e.g., cylindrical), with a comparable confining length (e.g., diameter). Thus, the issue of experimental observation of this predicted superfluid response of \paraH2 is centred on creating an adsorbing porous matrix, whose microscopic structure consist of interconnected nanoscale size pores, through which superflow could be established, at a temperature of the order of a fraction of a K. \\ \indent
While coating the walls of a porous glass with a relatively thick (a few layers) film of Cs or Rb, for example, may not be experimentally feasible at this time, perhaps techniques such as those utilized to fabricate nanohole semiconductor membranes, presently utilized to investigated hydrodynamics of superfluid helium in quasi-one-dimension \cite{gervais} could be adopted. 
\begin{figure}  [t]           
\centerline{\includegraphics[scale=0.36]{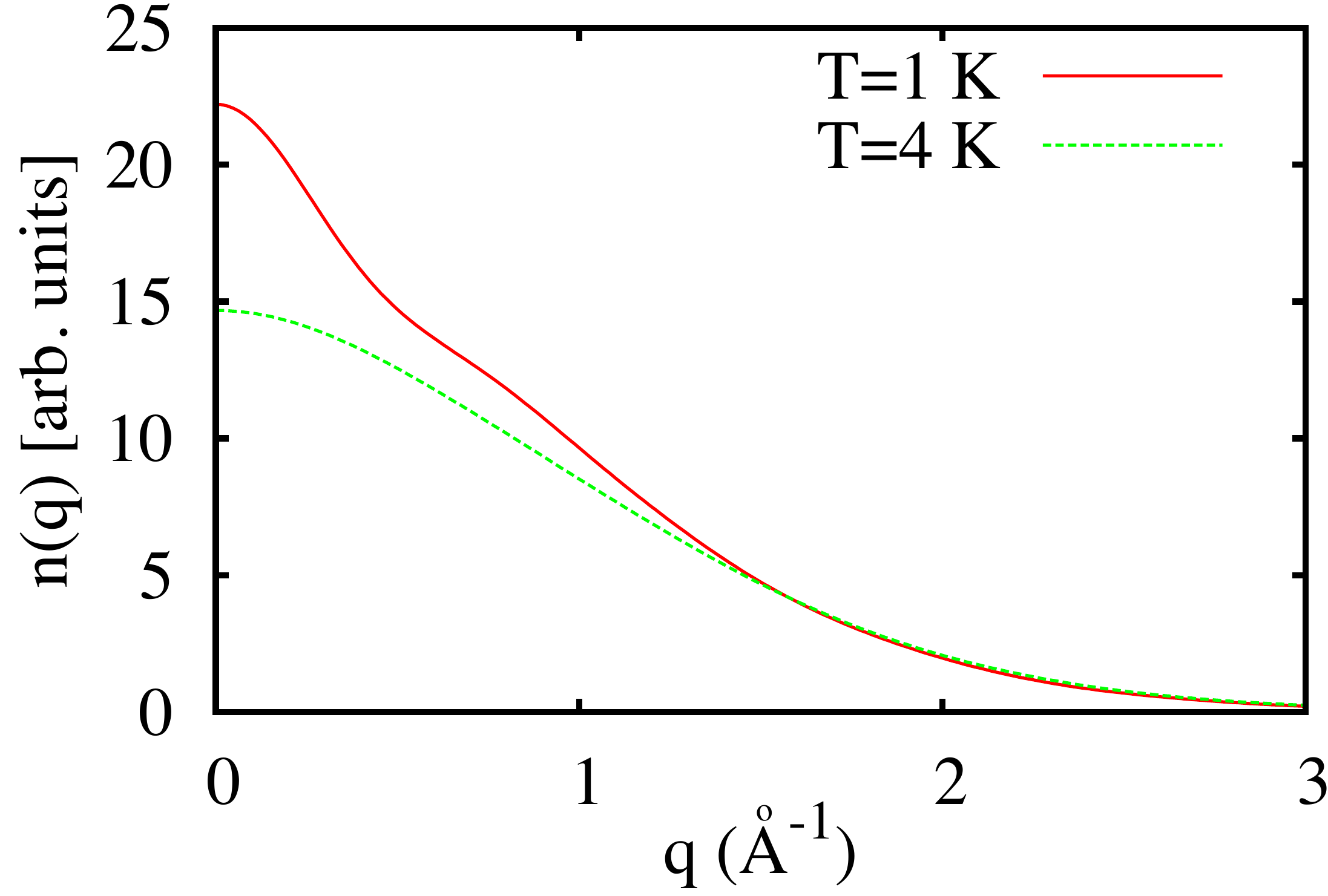}} 
\caption{Spherically averaged momentum distribution for a system of 32 \paraH2 molecules inside a  Cs cavity of radius 10 \AA\ at $T$=4 K and $T$=1 K.}
\label{nq}
\end{figure} 
\\ \indent
As a final remark, we mention that one might also think of detecting a superfluid transition of the embedded \paraH2 fluid  by measuring the momentum distribution, typically by neutron scattering, and looking for the appearance at low temperature of a peak at zero momentum,\cite{sokol} which signals the onset of Bose-Einsten condensation, intimately connected to superfluidity. This requires no actual bulk superflow, nor the use of molecular probes, which renders the unambiguous detection of superfluidity problematic in free standing clusters.\cite{grebenev} While the peak is not sharp as it would be in bulk superfluid, but rather broadened by the fact the system is confined over a length of $\sim$ 1 nm, its detection should still be possible (Fig. \ref{nq}). In contrast, no such a peak develops if crystallization occurs inside the cavities, as observed experimentally.\cite{sokol}
\\ \indent
This work was supported by the Natural Sciences and Engineering Research Council of Canada (NSERC).  M. B. gratefully acknowledges the hospitality of the Max-Planck Institute for the Physics of Complex Systems in Dresden, and of the Theoretical Physics Institute of the ETH, Z\"urich, where part of this work was carried out.

\end{document}